\documentclass[11pt]{article}
\usepackage{amsmath}
\usepackage{graphicx}
\usepackage{amssymb}
\usepackage{amsfonts}
\usepackage{latexsym}
\usepackage{psfig}

\def\beq{\begin{eqnarray}}
\def\eeq{\end{eqnarray}}
\def\ln{\,\mbox{ln}\,}

\def\tr{\,\mbox{tr}\,}
\def\Tr{\,\mbox{Tr}\,}

\def\al{\alpha}
\def\be{\beta}

\def\ga{\gamma}

\def\ep{\epsilon}

\def\la{\lambda}
\def\na{\nabla}

\def\om{\omega}
\def\ph{\varphi}
\def\ta{\tau}

\def\Ga{\Gamma}

\begin{document}

\begin{center}
{\Large\sc On the renormalization group for the interacting
\\
massive scalar field theory in curved space}
\vskip 6mm

{\small \bf 
Guilherme de Berredo-Peixoto $^{(a)}$}
 \footnote{E-mail address: guilherme@fisica.ufjf.br},
\quad 
{\small \bf 
Eduard V. Gorbar $^{(b)}$}
\footnote{E-mail address: gorbar@bitp.kiev.ua},
\quad 
{\small \bf Ilya L. Shapiro$^{(a)}$}
\footnote{On leave from Tomsk State Pedagogical University,
Russia. E-mail address: shapiro@fisica.ufjf.br} 

\vskip 6mm

{\small\sl
(a) \quad 
Departamento de F\'{\i}sica -- ICE,
Universidade Federal de Juiz de Fora,} \\
{\small\sl 
Juiz de Fora, CEP: 36036-330, MG,  Brazil
\vskip 4mm

(b) \quad 
Bogolyubov Institute for Theoretical Physics, Kiev, Ukraine}

\vskip 2mm
\end{center}

\vskip 6mm


\begin{quotation}
\noindent
{\large\it Abstract.}
The effective action for the interacting massive scalar field 
in curved space-time is derived using the heat-kernel method. 
Starting from this effective action, we establish a smooth 
quadratic form of the low-energy decoupling for the four-scalar 
coupling constant $\,\la\,$ and for the nonminimal interaction 
parameter $\,\xi$. The evolution of this parameter from the 
conformal value $\,1/6\,$ at high energies down to the IR regime 
is investigated within the two toy models with negative and  
positive four-scalar coupling constants. 
\vskip 3mm

PACS: $\,$ 04.62.+v 
$\,\,$
11.10.Hi  
$\,\,$
11.10.Jj
\vskip 3mm

\end{quotation}

\vskip 12mm

\section{\large\bf Introduction}

\qquad
The relevance of scalar fields is significant in both particle
physics and cosmology. For instance, the Higgs scalar is in 
the heart of the Standard Model of particle physics (SM) 
providing the possibility of the Spontaneous Symmetry Breaking 
and the Higgs mechanism. Furthermore, many extensions of 
the SM imply new scalars corresponding either to the new 
symmetries (e.g. supersymmetry) or to their breaking. In 
cosmology
the scalar fields are invoked to mimic the variable vacuum 
energy. In particular, they represent an important element 
of the inflaton models and also, as different sorts of 
quintessence, may play the role of the Dark Energy in the late
Universe. Obviously, the cosmological applications of scalar 
fields indicate the special 
importance of their interaction to gravity. It is well-known 
that there is an arbitrary element called a non-minimal term. 
Say, the Lagrangian of a single scalar $\,\ph\,$ includes, 
along with the 
covariant kinetic and massive terms $\,(\na \ph)^2+m^2\ph^2$, 
the non-minimal term $\,\xi R\ph^2$, where $\,R\,$ is the 
scalar curvature and
$\,\xi\,$ is the parameter of the non-minimal interaction. 
This term plays a special role, because it represents the 
unique possible non-minimal structure with the dimensionless 
coefficient. All other terms include coefficients with the 
inverse-mass dimensions and therefore will be strongly 
suppressed at low energies. In this situation, any information 
concerning the value of $\,\xi\,$ would be significant for 
the applications.

In classical theory the value of $\,\xi\,$ is 
arbitrary, but at the quantum level one can impose some 
constraints. In the massless case $\,m=0$, the value 
$\,\xi=1/6\,$ corresponds to 
the theory with the local conformal symmetry. Only in this
case the trace of the Energy-Momentum Tensor for the scalar 
field equals zero. On the other hand, we know that the  
Energy-Momentum Tensor of a massless particle 
has zero trace. Therefore, the value $\,\xi=1/6\,$ is the
only choice which provides a correspondence between field
and particle descriptions in the massless limit (see, 
e.g. \cite{wave}).
The massless limit means, in particular, that the equation 
of state for the ideal gas of the corresponding particles 
is $\,\rho=3p$. Of course, any 
gas of identical massive particles will approach this
equation of state when the kinetic energies of these 
particles become much greater that their masses. Therefore, 
we can suppose that the UV limit for the scalar field
corresponds to $\,\xi=1/6$. Then, in order to learn 
the value of $\,\xi\,$ at the lower energies we have 
to start from the conformal point $\,\xi=1/6\,$ in the UV. 
Furthermore, Quantum Field Theory provides a natural 
mechanism for calculating the values of $\,\xi\,$ at the
lower energies through the Renormalization Group 
equations formulated in curved space-time 
\cite{nelpan,tmf,book}. In the early works 
\cite{nelpan,tmf} on the subject (see the review and many 
other references in \cite{book}) the main 
attention has been paid to the renormalization group 
running of $\,\xi\,$ in various gauge models. 
The remarkable achievements were the discovery 
of the models with the UV stable conformal fixed point
\cite{buod} and the systematic investigation of the
possibility of such fixed point \cite{buya}. The
next step has been done in \cite{brv}, where the problem
of how the system evolves {\sl starting from} the 
conformal fixed point has been carefully studied, 
taking into account the possible quantum effects of the 
conformal factor \cite{antmot} of the metric and the 
higher-loop effects. Another framework for taking the 
quantum gravity effects into account is the higher derivative 
quantum gravity \cite{bush}, where the renormalization 
group equation for $\,\xi\,$ gains additional contributions
from the gravitational field loops.

The common point of all mentioned approaches is that they
are based on the most simple Minimal Subtraction scheme
of renormalization. On the other hand, despite this scheme 
is very efficient in the UV corner of the theory, it is 
not really trustable when we intend tracing back the running
from the massless UV fixed point to the intermediate or
even IR regime. In this case the masses of the particles 
become important, because they modify the Renormalization 
Group equations (RG). An alternative approach, based on 
the more physical mass-dependent renormalization scheme
(see, e.g. \cite{manohar} for the introduction and
further references),
has been applied to the gravitational problems just 
recently \cite{apco,fervi}. The focus of attention
of these works was the renormalization group and decoupling
in the vacuum sector of the quantum field theory in an 
external classical
gravitational background. It turned out, that the 
physical renormalization group can be formulated in the 
framework of the linearized gravity. The low-energy 
decoupling of the massive scalar \cite{apco}, fermion
and vector fields (including the QCD constituents) \cite{fervi}
performs smoothly, similar to the Appelquist and Carazzone
theorem in QED \cite{AC}. In the present letter we 
apply the same method to the analysis of the 
renormalization group for the non-minimal parameter $\,\xi$.
For this end we consider the theory with the action 
\beq
S = \int\,d^4x\sqrt{|g|}\,\left\{\,
\frac{1}{2}\,\na_\mu\phi\na^\mu\phi 
- \frac{1}{2}\,m^2\phi^2 + \frac{1}{2}\,\xi R\phi^2-
\frac{\la}{24}\phi^4\,\right\} 
\label{action}
\eeq
and derive both divergent and finite parts of the 
one-loop effective action in the matter field sector.
In principle, such calculation can be performed using 
the Feynman diagrams \cite{apco}, but since the heat-kernel 
technique \cite{avramidi,bavi2} provides the possibility
to perform the calculation in the most economic way,
we shall use this method in the way adapted in 
\cite{apco,fervi}.

The paper is organized as follows. In the next section 
we derive the one-loop effective action. In section 
3 we obtain the $\be$-functions for the coupling $\,\la\,$
and the 
non-minimal parameter $\,\xi$, and establish the explicit
form of their decoupling at low energies. Furthermore, we
discuss the scale dependence of $\,\la\,$ and $\,\xi\,$. 
In section 4 we draw our conclusions and outline the 
prospects for the future work.

\section{\large\bf Derivation of Effective Action}

\qquad
The one-loop Euclidean 
effective action corresponding to the theory (\ref{action})
is given by 
\beq
\bar{\Ga}^{(1)}\,=\,-\,\frac{1}{2}\,{\rm Tr\, ln} \,\left[\,
-\hat{1}\,\Box - \hat{1}\,m^2 - \hat{P} + \frac{\hat{1}}{6}\,R
\,\right]\,, 
\label{trln}
\eeq
where the hats indicate operators acting in the space of the 
quantized fields. As far as we are going to 
consider only the scalar field $\,\ph$, for us $\,\hat{1}=1$, 
where $\,1\,$ is the image of the delta function. Furthermore,
$\,\hat{P} = -(\xi -1/6)R + \la\phi^2/2$. The effective action 
(\ref{trln}) can be expressed by means of the proper time 
integration of the heat kernel
\beq
\bar{\Ga}^{(1)} = -\frac{1}{2}\int_0^\infty\,\frac{ds}{s}
\,\Tr K(s)\,.
\label{heat}
\eeq
As usual, the last expression can be expanded into the powers 
of the field strengths (curvatures), $\,R_{\mu\nu\al\be}\,$ 
and $\,\hat{P}$.
We would like to remind that the one-loop effective
potential contains arbitrary powers of the field $\phi$ 
but (by definition) not its derivatives. In the present 
case we expand the effective action up to the fourth order 
in $\phi$ but take into account an arbitrary number of 
derivatives. The derivative expansion of the one-loop 
effective action with an arbitrary powers of $\,\phi\,$ 
has been considered earlier in \cite{GK}.

 Up to the second order in curvatures, the 
expansion has the form \cite{avramidi,bavi2}
\beq
\Tr K(s) & = & \frac{\mu^{4-2\om}}{(4\pi s)^\om}
\,\int d^{2\om -4}x\,\sqrt{g}
\,\,e^{-sm^2}\tr \left\{\, \hat{1}+ s\hat{P}
\,+\, s^2\,\big[\, 
R_{\mu\nu}f_1(-s\Box)R^{\mu\nu}+ \right. 
\nonumber 
\\
& + & \left. Rf_2(-s\Box)R + \hat{P}f_3(-s\Box)R + 
\hat{P}f_4(-s\Box)\hat{P}
\big]\,\right\}\, .
\label{heat 2}
\eeq
Here $\om$ is the dimensional parameter, $\mu$ is an arbitrary
renormalization parameter 
with dimension of mass and the functions $f_i$ are given by
\beq
f_1(\ta) = \frac{f(\ta)-1+\ta /6}{\ta^2}\, ,\;\;\;\;\;\;
f_2(\ta) = \frac{f(\ta)}{288}+\frac{f(\ta)-1}{24\ta}-
\frac{f(\ta)-1+\ta /6}{8\ta^2}\, ,
\nonumber
\eeq
\beq
f_3 = \frac{f(\ta)}{12}+\frac{f(\ta)-1}{2\ta}\, , \;\;\;\;\;\;
f_4 = \frac{f(\ta)}{2}\, ,
\qquad\mbox{where}\qquad
f(\ta) = \int_0^1 d\al\, e^{\al (1-\al)\ta}\, , \;\;\;\;\; \ta = -s\Box\, .
\nonumber
\eeq
After straightforward calculations \cite{fervi}, using
the notations $\,t=sm^2\;,\quad u=\ta /t =-\Box /m^2$,
we obtain an explicit expression for the Effective Action
up to the second order in $\,{\hat P}\,$ and $\,R_{\mu\nu}$ 
\beq
\bar{\Ga}^{(1)}\,=\, -\frac{1}{2(4\pi)^2}
\,\int d^{2\om -4}x\sqrt{g}\,
\left(\frac{m^2}{4\pi\mu^2}\right)^{\om-2}\int_0^\infty\, 
dt e^{-t}\,\Big\{\, \frac{m^4}{t^{\om +1}} 
- \frac{m^2}{t^\om}\left(\xi -\frac{1}{6}\right) R 
\nonumber 
\\
+\frac{m^2}{4t^\om}\la\phi^2 
+ \sum_{i=1}^{5}l^*_iR_{\mu\nu}M_iR^{\mu\nu} 
+ \sum_{j=1}^{5}l_jRM_jR 
+ \sum_{k=1}^{5}l^{{\xi}}_k\,R\,M_k\,\phi^2 
+ \sum_{n=1}^{5}l^{\la} _n\phi^2\,M_n\,\phi^2\,\Big\}\, ,
\label{EA}
\eeq
where (we correct a misprint of \cite{fervi} in the expression
for $M_1$) 
\beq
M_1 = \frac{f(tu)}{t^{\om -1}}\, , \;\;\; M_2 
= \frac{f(tu)}{t^\om u}\, , \;\;\;
M_3 = \frac{f(tu)}{t^{\om +1}u^2}\, ,\;\;\; M_4 
= \frac{1}{t^\om u}\, ,\;\;\; 
M_5 = \frac{1}{t^{\om +1}u^2}\, , \nonumber
\eeq
and the coefficients have the form
\beq
l^*_{1,2} = 0\, ,\;\;\;\;\; l^*_3 = 1\, , 
\;\;\;\;\; l^*_4 = \frac{1}{6} 
\, , \;\;\;\;\; l^*_5 = -1\, , \nonumber
\eeq
\beq
l_1 = \frac{1}{288} - \frac{1}{12}\left(\xi-\frac{1}{6}\right)+
\frac{1}{2}\left(\xi-\frac{1}{6}\right)^2\, , \;\;\;\;\;\;
l_2 = \frac{1}{24} - \frac{1}{2}\left(\xi-\frac{1}{6}\right)\, , 
\nonumber
\eeq
\beq
l_3 = -\frac{1}{8}\, , \;\;\;\;\;\; l_4 = -\frac{1}{16}+
\frac{1}{2}\left(\xi-\frac{1}{6}\right)\, ,
\;\;\;\;\;\; l_5 = \frac{1}{8}
\, , \nonumber
\eeq
\beq
l^{{\xi}}_1 = -\frac{\la}{2}\left(\xi-\frac{1}{6}\right)
+ \frac{\la}{24}\, , \;\;\;\;
l^{{\xi}}_2 = \frac{\la}{4}\, , 
\;\;\;\; l^{{\xi}}_{3,5} = 0\, , \;\;\;
l^{{\xi}}_4 = -\frac{\la}{4}\,, 
\nonumber
\eeq
\beq
l^{\la}_1 = \frac{\la^2}{8}\,,\;\;\;\;
l^{\la}_{2,3,4,5} = 0\,. 
\nonumber
\eeq
Indeed, the coefficients $\,l_i\,$ and $\,l^*_j\,$ are 
the same as for the free scalar field \cite{apco}, while 
the last two sets are completely due to the interaction
and do not have analogs in the free field case.
At this point one has to perform the integration in the 
variables $\,t\,$ and $\,\al$. The result can be expressed 
in terms of notations \cite{apco}
\beq
A\,=\,1-\frac{1}{a}\ln\frac{1+a/2}{1-a/2}\,,
\qquad a^2 = \frac{4\Box}{\Box - 4m^2}\,. 
\label{Aa}
\eeq
The calculation of the integrals needs the expansions
\beq
\left(\frac{m^2}{4\pi\mu^2}\right)^{\om -2} 
= 1+(2-\om )\ln\left(\frac{4\pi\mu^2}{m^2}\right) + ...
\label{integrals2}
\eeq
and
\beq
\Ga (2-\om ) & = & \frac{1}{2-w} - \ga +{\cal O}( 2-\om )\,,
\nonumber \\
\Ga (1-\om ) & = & - \frac{1}{2-w} + \ga
 -1+ {\cal O}( 2-\om )\, ,\nonumber \\
\Ga (-\om )\quad & = & \frac{1}{2(2-w)} - \frac{\ga}{2} 
+ \frac{3}{4}+ {\cal O}( 2-\om )\,.
\nonumber
\eeq

The integrals of expressions $\,e^{-t}M_i\,$ 
are listed below:
\beq
\int_0^\infty dt e^{-t}\, M_1 =  \frac{1}{\ep} 
+ 2A  +{\cal O}(2-\om )\, ,\nonumber
\eeq
\beq
\int_0^\infty dt e^{-t}\, M_2 
= \left(\frac{1}{12}-\frac{1}{a^2}\right)\,\left(
\frac{1}{\ep} + 1 \right) - \frac{4A}{3a^2}
+ \frac{1}{18} + {\cal O}(2-\om)\, , \nonumber
\eeq
\beq
\int_0^\infty dt e^{-t}\, M_3 = \left(\frac{1}{2a^4}
- \frac{1}{12a^2}+\frac{1}{60}\right)
\left(\frac{1}{\ep}  +\frac32\right) 
+ \frac{8A}{15a^4} - \frac{7}{180a^2} + \frac{1}{400} + 
{\cal O}(2-\om)\, , \nonumber
\eeq
\beq
\int_0^\infty dt e^{-t}\, M_4 = \frac{a^2-4}{4\,a^2}
\,\left(\,\frac{1}{\ep} + 1\right) + {\cal O}(2-\om )\, , 
\nonumber
\eeq
\beq
\int_0^\infty dt e^{-t}\, M_5 
= \frac{(a^2-4)^2}{32\,a^4}\,\left(
\frac{1}{\ep}+\frac32 \right) + {\cal O}(2-\om )\,,
\label{M}
\eeq
where we denoted
$$
\frac{1}{\ep}=\frac{1}{2-w}
+\ln \Big(\frac{4\pi \mu^2}{m^2}\Big) - \ga\,.
$$

Let us notice that, different from the previous publications
\cite{apco,fervi}, we included the Euler constant into 
(\ref{M}) explicitly. Indeed, this does not 
change the final result for the effective action or for the
$\,\be$-functions, because this constant can be always 
removed by the change of the renormalization parameter 
$\,\mu$. After some algebra, we arrive at the expression 
for the one-loop effective action
$$
{\bar \Ga}^{(1)}_{scalar}
\,=\,\frac{1}{2(4\pi)^2}\,\int d^4x \,g^{1/2}\,
\left\{\,\frac{m^4}{2}\cdot\Big(\frac{1}{\ep}
+\frac32\Big)\,+
\,\Big(\xi-\frac16\Big)\,m^2R\,
\Big(\frac{1}{\ep}+1\Big)
\right.
$$$$
\left.
+\,\frac12\,C_{\mu\nu\al\be} 
\,\Big[\frac{1}{60\,\ep}+k_W(a)\Big] C^{\mu\nu\al\be}
\,+\,R \,\Big[\,\frac{1}{2\ep}\,\Big(\xi-\frac16\Big)^2\,
+ k_R(a)\,\Big]\,R
\right.
$$
\beq
\left.
-\,\frac{\la}{2\ep}\,m^2\phi^2 
\,+ \,\phi^2\Big[\frac{\la^2}{8\ep}+k_\la(a)\Big]\phi^2
\,+ \,\phi^2\Big[
-\frac{\la}{2\ep}\Big(\xi-\frac16\Big)+k_\xi(a)\Big]\,R
\,\right\}\,.
\label{final}
\eeq
The expressions for 
$\,k_W(a)\,$ and $\,k_R(a)\,$ can be found in \cite{fervi} 
and the new formfactors corresponding to the scalar 
self-interaction and non-minimal interaction between 
scalar and metric are 
\beq
k_\la(a) \,=\,\frac{A\, \la^2}{4}\,,
\label{form lambda}
\eeq
\beq
k_\xi(a) \,=\,\la\left[\,\frac{A\,(a^2-4)}{12\,a^2}
\,-\,\frac{1}{36}\,-\,A\,\Big( \xi - \frac16\Big)\,\right]\,,
\label{form xi}
\eeq
where we used the notations (\ref{Aa}).
The two formfactors (\ref{form lambda}), (\ref{form xi}) 
contain all information about the scale dependence of 
the parameters $\,\la\,$ and $\,\xi$. In the next section 
we shall discuss this dependence in details.

\section{\large\bf 
Renormalization group equations and running parameters}

\qquad
In the modified Minimal Subtraction scheme ($\overline{{\rm MS}}$)
 the $\be$-function of the effective 
charge, $C$, is defined as 
\beq
\be_C(\overline{{\rm MS}}) 
= \lim_{n\to 4}\,\mu\,\frac{dC}{d\mu}\, .
\eeq
When we apply this procedure to the expression (\ref{final}) 
the $\be$-functions of $\la$ and $\xi$, in the 
$\overline{{\rm MS}}$ scheme, coincide with the usual ones,
obtained in the local covariant approach \cite{book}. 
The main advantage of the mass-dependent scheme is that 
the renormalization point where the subtraction 
is done ($p^2 = M^2$) has a particular physical meaning, 
giving us explicit information on the decoupling
of massive fields in external gravitational background.

The $\be$-function is defined, in the mass-dependent scheme, 
as $\,\lim\limits_{n\to 4}\,M\frac{d}{dM}\,$ applied to the 
regularized form-factor of the corresponding term in the 
effective action.
This procedure is equivalent to taking $\,-pd/dp\,$ 
of the formfactor of the $\,\phi^4\,$ and $\,R\phi^2\,$-terms. 
The $\,\be_\la$-function which follows from this procedure has 
the form
\beq
\be_\la = \frac{3\,\la^2}{4(4\pi)^2}\,\big(a^2-a^2A+4A\big)\,.
\label{beta lambda}
\eeq
The UV limit corresponds to $\,a\to 2\,$ and gives 
\beq
\be_\la^{UV} = \frac{3\,\la^2}{(4\pi)^2}\,,
\label{beta lambda UV}
\eeq
providing a perfect fit with the well-known result of the
$\overline{\rm MS}$-scheme 
$\,\be_\la^{UV}=\,\be^{\overline{\rm MS}}_\la\,$. 
In the IR limit $\,a\to 0\,$ 
we meet 
\beq
\be_\la^{IR} = \frac{\la^2}{2\,(4\pi)^2}\,\cdot\,\frac{p^2}{m^2}
\,+\,{\cal O} \Big( \frac{p^4}{m^4} \Big)\,.
\label{beta lambda IR}
\eeq
Clearly, this means the standard IR decoupling for the effective
charge $\,\la$, similar to the Appelquist and Carazzone theorem
in QED \cite{AC}.

Let us use the $\,\be$-function (\ref{beta lambda}) to 
investigate the running of $\,\la\,$ between the UV and 
IR limits\footnote{We would like to point out that the 
$\beta$-function 
for $\,m^2\,$ is exactly zero at one loop.}. The solution 
of the renormalization group equation 
\beq
\mu\frac{d\la(\mu)}{d\mu}\,=\,\be^{\overline{\rm MS}}_\la\,,
\qquad \la(m)=\la
\label{RG lambda MS}
\eeq
is well-known
\beq
\lambda(\mu)\,=\,\frac{\la}{1-\frac{3\la}{(4\pi)^2}\,\ln(\mu/m)}\,.
\label{solution lambda MS}
\eeq
The expression above is singular in the UV where it manifests
the Landau pole for $\,\la>0\,$ and also singular in the IR 
for  $\,\la<0$. The last property is because the 
$\overline{\rm MS}$-scheme is not sensitive to the presence
of the mass of the field. 

It is remarkable that the renormalization group equation 
with the physical $\,\be_\la$-function 
\beq
p\frac{d\la(p)}{dp}\,=\,\be_\la(\la,m,p)\,,
\qquad \la(m)=\la
\label{RG lambda}
\eeq
also admits analytic solution in terms of elementary 
functions
\beq
\la(p)\,=
\,{\la}\,\left[1 - \frac{3\,\la}{(4\pi)^2}\,
\Big(\sqrt{5}\ln \frac{\sqrt{5}-1}{\sqrt{5}+1}  
\,+\,\frac{\sqrt{4m^2 + p^2}}{p}\,
\ln \frac{\sqrt{4m^2+p^2}+p}{\sqrt{4m^2 + p^2}-p}
\Big)\right]^{-1}\,.
\label{solution lambda}
\eeq
Qualitatively, the behaviours of the two functions 
(\ref{solution lambda MS}) and (\ref{solution lambda}) are 
not very different. In particular, both manifest the Landau 
pole, as can be seen at the Figure 1
\footnote{All the plots have been obtained
using the Mathematica program \cite{M4}.}. 
The main difference is the position of 
this pole which is moving farther into UV limit for the 
case of the physical renormalization group 
(\ref{solution lambda}). Of course, from the physical point 
of view this difference is not very important, at least when 
we have in mind the Higgs field. The reason
is that the Landau pole corresponds to extremely high
energies. For example, if we assume that the scalar mass
is $100\,GeV$ and that $\,\la=1$ (extreme case, for smaller
values of $\,\la\,$ the pole moves even to much higher 
energies), the 
position of the pole is at $p\approx 10^{23}\,GeV$. This 
energy range is far beyond the Planck scale, and therefore
it does not pose a problem for, e.g. the Standard Model. 
However, it makes a problem for us, because our main  
purpose is to evaluate a running of the parameter $\,\xi\,$
from the conformal UV point $\,\xi=1/6\,$ down to the 
IR regime. In this situation we decided to use the following 
two toy models: 
one with $\,\la\equiv 1$ (let us notice that the greater
value of $\,\la\,$ enforces the running of $\,\xi$) and  
another one with the negative sign of the coupling $\,\la$. 
The main disadvantage of the last model is that the 
classical potential 
of this theory is not bounded from below. However, there 
is a serious advantage - the asymptotically free (AF) 
behaviour in the UV. Let 
us emphasize that we consider the theory with $\,\la<0\,$
just as a toy model which is called to mimic the 
renormalization group running of $\,\xi\,$ in a general 
AF GUT-like theory. The derivation of the physical 
$\be$-functions for the general GUT theory is much 
more involved and will be reported elsewhere. For the 
negative initial value of $\,\la\,$ the behaviour of the
corresponding effective charge is presented at the Figure 2.


\begin{figure}
\centerline{\psfig{file=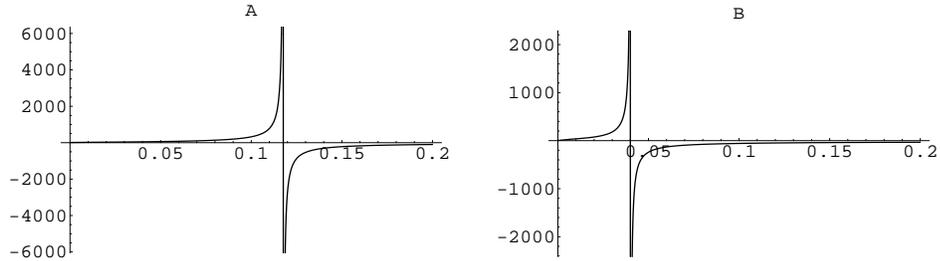,width=13cm,height=5cm}}
\vspace*{3pt}
\caption{{\it Plots for the scale dependence for the effective 
charge ({\sl A)} $\,\lambda(p/m)$ in the mass-dependent scheme 
of renormalization, and ({\sl B)} $\,\lambda(\mu/m)$ 
in the $\overline{\rm MS}$-scheme. 
Both plots correspond to the initial data $\,\lambda(\mu=m)=0.1\,$ 
and the units $\,10^{230}\,$ for both $\,\mu/m\,$ and $\,p/m$.}}
\end{figure}
 

\begin{figure}
\centerline{\psfig{file=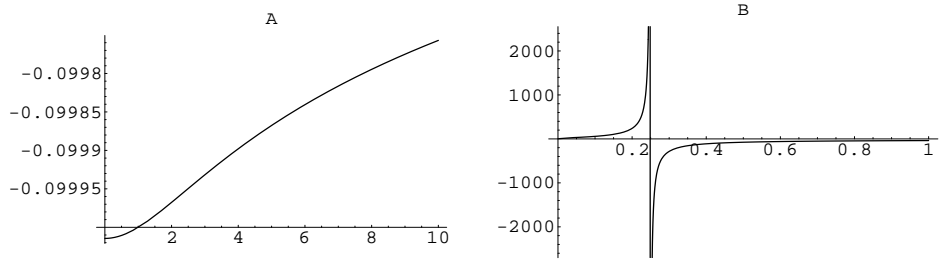,width=13cm,height=5cm}}
\vspace*{3pt}
\caption{{\it The plots representing the scale 
dependence of $\,\la\,$ for the AF model with $\,\lambda < 0$.
Two cases are presented: \quad {\sl (A)} $\,\lambda\,$ as 
a function of $\,p/m\,$ 
for the mass-dependent scheme of renormalization);
\quad {\sl (B)} $\,\lambda\,$ as a function of $\mu/m$ for 
the $\,\overline{\rm MS}$-scheme. In the last case the units of 
$\,\mu/m\,$ are $10^{-228}$.
It is remarkable that the plot {\sl (A)} shows regular behaviour 
in the IR, while the plot {\sl (B)} has an IR pole, like 
for the massless theory. For $p>m$ the two plots are almost 
identical.}}
\end{figure}


For the $\,\be_\xi$-function we arrive at the following 
result:
\beq
\be_{\xi}\,=\,\frac{\lambda}{48\,(4\pi)^2\,a^2}\,\Big[
(a^2-4)(a^2\,A-12\,A-a^2)
\,+\, 12a^2\Big(4\,A+a^2-a^2\,A \Big)\,\Big(\xi-\frac16\Big)
\Big]\,.
\label{beta xi}
\eeq
It is easy to see that the last expression meets standard 
tests and expectations. In the UV limit $\,a\to 2\,$ there is 
correspondence with the known $\overline{\rm MS}$-scheme result
\beq
\be^{UV}_{\xi}\,=\,\be^{\overline{\rm MS}}_{\xi}\,=\,
\frac{\la}{(4\pi)^2}\,\Big(\xi-\frac16\Big)\,.
\label{beta xi UV}
\eeq
In the low-energy regime there is a standard quadratic 
decoupling of the massive field
\beq
\be^{IR}_{\xi}
\,=\,\frac{\la}{6\,(4\pi)^2}
\,\left[\,\Big(\xi-\frac16\Big) - \frac{1}{30}\,\right]
\cdot \frac{p^2}{m^2} \,+\,{\cal O}\Big(\frac{p^4}{m^4}\Big)\,.
\label{beta xi IR}
\eeq
Thus, we have obtained the analog of the Appelquist and 
Carazzone theorem in the matter field sector for the 
theory of a massive scalar field in curved 
space-time. The decoupling takes place not only for the 
four-scalar coupling $\,\la$, but also for the nonminimal
parameter $\,\xi$.

Let us pay attention to the general form of the 
expression for the $\,\be_\xi$-function (\ref{beta xi}). 
In the UV limit, 
this $\,\be$-function contains the factor $\,(\xi-1/6)$. 
This property of the 
$\,\be_\xi$-function is not accidental, it is related 
to the conformal invariance of the one-loop divergences 
which can be proved in a general form \cite{tmf,book}. 
At the same time, out of the UV limit the 
$\,\be_\xi$-function is not proportional to $\,(\xi-1/6)$
due to the effect of the mass of the field. This is, 
indeed, the most important difference between  
$\,\be_\xi\,$ and $\,\be^{\overline{\rm MS}}_\xi\,$.

Using the momentum-subtraction scheme of renormalization,
in the framework of linearized gravity \cite{apco} one
can consider the renormalization group equation for 
the effective charge $\,\xi$
\beq
p\frac{d\xi(p)}{dp}\,=\,\be_\xi(\la,m,p)\,,
\qquad \xi(p_0)=\xi\,,
\label{RG xi}
\eeq
where the $\,\be_\xi\,=\,\be_\xi(\la,m,p)\,$-function is 
given by the expression (\ref{beta xi}). The last equation 
describes the reaction of the effective charge $\,\xi\,$ 
to the change of the momenta in the external loop of the 
corresponding Feynman diagram, shown at the Figure 3.
For the sake of completeness, we have also presented  
a diagram which contributes to the $\,\be_\la$-function. 


\begin{figure}
\centerline{\psfig{file=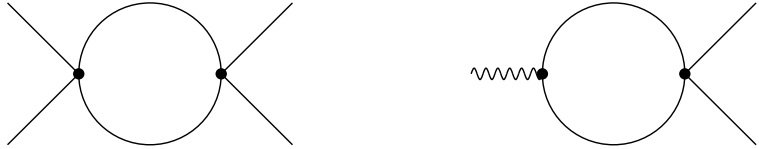,width=10.0cm,height=2cm}}
\vspace*{3pt}
\caption{{\it The relevant $1$-loop diagrams in the $\,\la\ph^4$-theory.
 The straight lines correspond to the massive scalar and 
 wavy lines to the external field $h_{\mu\nu}$. The 
 formfactor (\ref{form lambda}) coming from the first 
 diagram does not depend on the presence of external 
 $h_{\mu\nu}$ lines, while for the formfactor 
 (\ref{form xi}) coming from the second diagram at 
 least one of these external lines is necessary.}}
\end{figure}


Let us analyse the renormalization group equation 
(\ref{RG xi}). It is easy to see that (\ref{RG xi}) 
is nothing but a linear ordinary differential 
equation and therefore its analytic solution can be
easily obtained in terms of integrals involving the
expression (\ref{solution lambda}). The formula 
which follows from this procedure is extremely 
bulky and hence it is useless for us. Therefore we 
turn to the qualitative and numerical considerations.

In the momentum-subtraction renormalization 
scheme the $\,\be$-function is not directly linked to the 
divergences (except in the UV limit where the correspondence 
with the $\overline{\rm MS}$-scheme holds). That is why the 
conformal value $\,\xi=1/6\,$ is not a fixed point for the
theory of massive scalar field. At the same time, for the 
reasons explained in the Introduction, we have
to assume the conformal value as the initial 
point of the renormalization group trajectory in the UV. 
Therefore the most natural option is to 
impose the renormalization condition $\,\xi(p_0)=1/6\,$ in 
extreme UV $p_0^2 \gg m^2$ and integrate it numerically 
until the intermediate $\,p^2 \sim m^2\,$ or IR 
$\,p^2 \ll m^2\,$ regimes. 


\begin{figure}
\centerline{\psfig{file=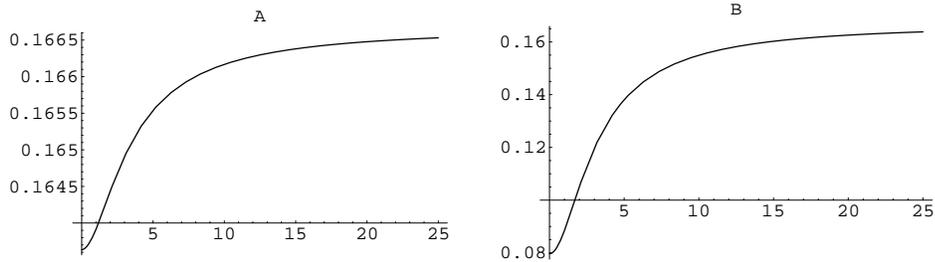,width=13.0cm,height=5.0cm}}
\vspace*{3pt}
\caption{{\it The plots of $\,\xi(p/m)$, corresponding to the initial 
value $\,\xi(p\gg m)=1/6\,$ and two 
different behaviours (A) $\,\la(p=m) = - 0.1\,$ 
(AF case with the unbounded from below potential) and 
(B) $\,\la \equiv 1$.}}
\end{figure}


The plots of the numerical solutions of the (\ref{RG xi}) 
for different behaviours of $\,\la=\la(p/m)\,$ are
presented at the Figure 4. As one can see at these plots,
the deviation of $\,\xi\,$ from the conformal value 
$\,1/6\,$ really takes place but the range of this 
deviation is not very big for the AF model with 
$\,\la(p=m) = - 0.1\,$. In the extreme case of large 
and constant coupling $\,\lambda\equiv 1$, the value 
of $\,\xi\,$ increases, approximately, $\,8\,\%\,$ 
between UV and IR. About the same range of running 
takes place between UV and the 
point $p^2=m^2$.  After all, despite the 
conformal value $\,\xi=1/6\,$ can not be exact in 
the massive scalar theory, it may serve as a reasonable 
approximation, at least in the UV region  
$\,p^2 > m^2$. The conformal approximation becomes 
very good for the AF theory with relatively weak 
coupling. 
Let us notice that the same conclusion 
has been obtained earlier in \cite{brv} within a very 
different approach, which involves IR quantum gravity
as a quantum theory of the conformal factor of the 
metric \cite{antmot}.

\section{Conclusions}

\qquad 
We have applied the exact solution for the heat-kernel 
of the second-order minimal operator \cite{avramidi,bavi2} 
for the 
derivation of the effective action for the interacting 
massive scalar field. The methods which were developed 
earlier in \cite{apco,fervi} made this calculation, 
technically, relatively simple and moreover open the 
possibility to derive the effective action of other 
interacting theories (e.g. Standard Model or GUT's) 
without using Feynman diagrams or, e.g. pinch technique. 
The last observation may be important, in particular,
for the investigations related to the supersymmetry
\cite{Brod}.

Using the effective action and the linearized gravity 
approach in curved space-time, we derived the renormalization 
group $\be$-functions for the four-scalar coupling constant 
$\,\la\,$ 
and for the nonminimal interaction parameter $\,\xi$. Both 
$\be$-functions correspond to the physical momentum-subtraction
scheme of renormalization and are essentially more complicated 
than the standard Minimal Subtraction $\be$-functions for the 
same effective charges. In the UV limit, however, there is a 
perfect correspondence between the two sets of the 
$\,\be$-functions. At low energies we observe the standard 
quadratic form of the decoupling for the scalar field, in 
accordance to the expectations based on the AC decoupling 
theorem \cite{AC}.

The scale dependence of the four-scalar coupling constant 
$\,\la\,$ does not depend on the presence of an external 
gravity field. Moreover, it is qualitatively the same as 
in the $\overline{MS}$ renormalization scheme. In both 
cases there is a Landau pole, but for the 
momentum-subtraction scheme case
the position of this pole is shifted (about 3 times) to the 
UV compared to $\overline{MS}$ case. If changing the sign
of the coupling $\,\la\,$, the theory becomes asymptotically 
free. Indeed, the theory with $\,\la<0\,$ is not a realistic 
model, because the scalar potential is not bounded from 
below. However, it may serve as a toy model for the 
investigation of the running of $\,\xi$. 
We traced the evolution of this parameter from the 
conformal value $\,1/6\,$ in the high emergy limit down to 
the IR regime and found that the numerical change of $\,\xi$
is not very big, in fact it does not exceed a  few percents
for the value $\,\la(p=m)=0.1$. One expect a similar range 
of running within the realistic AF models like GUT's.
\vskip 8mm

{\large\bf Acknowledgments.} 
The work of the authors has been partially supported by the 
research 
grant from FAPEMIG and by the fellowships from FAPEMIG 
(G.B.P.) and CNPq (I.Sh.). E.G. thanks Departamento 
de F\'{\i}sica at the Universidade Federal de Juiz de Fora
for kind hospitality. I.Sh. is grateful to the Erwin 
Schr\"odinger Institute for Mathematical Physics 
in Vienna and to the Departamento de F\'{\i}sica Te\'orica 
de Universidad de Zaragoza for kind hospitality and partial 
financial support.




\begin{thebibliography}{9}


\bibitem{wave} J.C. Fabris, A.M. Pelinson and I.L. Shapiro,
Nucl. Phys. {\bf B597} (2001) 539.

\bibitem{nelpan} B.L. Nelson and P. Panangaden, 
Phys. Rev. {\bf 25D} (1982) 1019; {\bf 29D} (1984) 2759.

\bibitem{tmf}  I.L. Buchbinder, 
Theor. Math. Phys. {\bf 61} (1984) 393.

\bibitem{book} I.L. Buchbinder, S.D. Odintsov, I.L. Shapiro,
{\sl Effective Action in Quantum Gravity} (IOP Publishing,
Bristol, 1992).

\bibitem{buod} I.L. Buchbinder and S.D. Odintsov,
 Sov. J. Phys. {\bf 25} (1983) 50.

\bibitem{buya} 
I.L. Buchbinder, I.L. Shapiro and E.G. Yagunov,
Mod. Phys. Lett. {\bf 5A} (1990) 1599.

\bibitem{brv} G. Cognola and I.L. Shapiro,
Class. Quant. Grav. {\bf 15} (1998) 3411.

\bibitem{antmot}{I. Antoniadis and E. Mottola,
Phys. Rev. {\bf 45D} (1992) 2013}.

\bibitem{bush} 
I.L. Buchbinder et al. Phys. Lett. {\bf 216B} (1989) 127;

I.L. Shapiro, Class. Quant. Grav. {\bf 6} (1989) 1197.

\bibitem{manohar} A. V.  Manohar, {\sl Effective Field
Theories} (Schladming 1996, Perturbative and nonperturbative
aspects of quantum field theory, [hep-ph/9606222]).

\bibitem{apco} 
E.V. Gorbar and I.L. Shapiro, JHEP {\bf 0302} (2003) 021.

\bibitem{fervi} 
E.V. Gorbar and I.L. Shapiro, JHEP {\bf 0306} (2003) 004;
{\sl Renormalization Group and Decoupling in Curved Space:
III. $\,$ The Case of Spontaneous Symmetry Breaking}, 
[hep-ph/0311190]. 


\bibitem{AC} T. Appelquist and J. Carazzone, Phys. Rev.
\textbf{11D} (1975) 2856.

\bibitem{avramidi} I. G. Avramidi, 
Yad. Fiz. (Sov. Journ. Nucl. Phys.) {\bf 49} (1989) 1185.

\bibitem{bavi2} A.O. Barvinsky and G.A. Vilkovisky, 
Nucl. Phys. {\bf B 333} (1990) 471.

\bibitem{GK} 
V.P. Gusynin and V.A. Kushnir, Sov. J. Nucl. Phys. 
{\bf 51} (1990) 373;  
Class. Quant. Grav. {\bf 8} (1991) 279.

\bibitem{M4} S. Wolfram, {\sl The Mathematica Book}
(Cambridge Univ. Press, 1999). 

\bibitem{Brod} M. Binger and S.J. Brodsky, 
{\sl Physical Renormalization Schemes and Grand Unification},
[hep-ph/0310322].

\end{thebibliography}
\end{document}